\documentclass[prl,aps,twocolumn,showpacs,superscriptaddress,floatfix,amsmath,amssymb,nofootinbib]{revtex4}

\usepackage[T1]{fontenc}
\usepackage{graphicx}
\usepackage{dcolumn}
\usepackage{bm}
\usepackage{array}
\usepackage{pdfpages}
\newcommand*{\defeq}{\mathrel{\vcenter{\baselineskip0.5ex \lineskiplimit0pt
                     \hbox{\scriptsize.}\hbox{\scriptsize.}}}
                     =}
\newcommand{\beq}{\begin{equation}}
\newcommand{\eeq}{\end{equation}}
\newcommand{\beqar}{\begin{eqnarray}}
\newcommand{\eeqar}{\end{eqnarray}}
\newcommand{\bea}{\begin{eqnarray}}
\newcommand{\eea}{\end{eqnarray}}
\newcommand{\bcen}{\begin{center}}
\newcommand{\ecen}{\end{center}}

\newcommand{\bra}[1]{\left< #1 \right|}
\newcommand{\ket}[1]{\left| #1 \right>}
\newcommand{\ketbra}[2]{\left| #1 \right> \left< #2 \right|}
\newcommand{\braket}[2]{\left< #1 \vert  #2 \right>}

\newcommand{\Mean}{\mathbf{M}}

\newcommand{\Hop}{\hat H}

\newcommand{\rop}{\hat\rho}

\newcommand{\sop}{\hat{\sigma}}

\newcommand{\pop}{\hat{p}}
\newcommand{\qop}{\hat{q}}

\begin{document}

\title{Action-Noise-Assisted Quantum Control}
\author{Amikam Levy}
\email{amikamlevy@gmail.com}
\affiliation{The Fritz Haber Research Center for Molecular Dynamics, The Institute  of Chemistry,  The Hebrew University, Jerusalem 91904, Israel}

\author{E. Torrontegui}
\email{eriktm@iff.csic.es}
\affiliation{The Fritz Haber Research Center for Molecular Dynamics, The Institute  of Chemistry,  The Hebrew University, Jerusalem 91904, Israel}
\affiliation{Instituto de F\'{\i}sica Fundamental IFF-CSIC, Calle Serrano 113b, 28006 Madrid, Spain}

\author{Ronnie Kosloff}
\affiliation{The Fritz Haber Research Center for Molecular Dynamics, The Institute  of Chemistry,  The Hebrew University, Jerusalem 91904, Israel}

\begin{abstract}
We study the effect of action noise on state-to-state control protocols.
Action noise creates dephasing in the instantaneous eigenbasis of the Hamiltonian and hampers the fidelity of the final state with respect to the target state. 
We find that for shorter protocols the noise more strongly influences the dynamics and degrades fidelity.
We suggest improving the fidelity by inducing stronger dephasing rates along the process. 
The effects of action noise on the dynamics and its manipulation is described for a general Hamiltonian and is then studied by examples.

\end{abstract}  
\date{\today}
\maketitle
\section{Introduction}
The aim of quantum control theories is to develop protocols to prepare entangled states, coherent states, or any other state possessing novel quantum properties \cite{torrontegui13,koch15,petersen10,rabitz00,lloyd99,vitanov_pulses,lidar}.
These methods are applicable in a wide variety of fields including 
quantum computation \cite{sarandy11}, cooling \cite{sta_prl1}, quantum transport \cite{erik_transport,bowler}, 
quantum state preparation \cite{muga12b, 2levelEXP1, 2levelEXP2}, cold atoms manipulation \cite{erik_bec,sta_exp1,schaff11,split_sta},
many-body state engineering \cite{adolfo_multi,SebCriAdo} as well as many
other applications in metrology, atomic, molecular and optical physics.
The main hindrance in manipulating quantum systems is the unavoidable presence of noise during the process.
\par
In this work we show that by inducing strong dephasing  we can improve the state-to-state controllability of a system. 
Generally, noise hampers the fidelity of the controlled state with respect to its target state \cite{khasin11,koch16}. 
In such cases, speeding up a process can suppress the influence of the noise by reducing the time during which the noise disturbs the system, allowing for high-fidelity execution of a protocol.
Shortcut to adiabaticity protocols (SP) \cite{torrontegui13} are examples of such methods in which during the process the state is not necessarily an eigenstate of the instantaneous time dependent Hamiltonian, but it becomes so in the final time. 

Here, we focus on SP in the presence of action noise.
The scheme introduced in this paper applies for initial states that are eigenstates of the  Hamiltonian and are diagonal in this basis, including mixed states such as thermal states.  
We show that the noise becomes more influential in short time operations, implying lower fidelity in the controled state.
We further show that by increasing the dephasing rate along the process, the fidelity is enhanced, meaning that above a certain threshold intensifying the noise becomes beneficial.
The dephasing rate can be controlled by manipulating the Hamiltonian during the process, 
implying that for a given initial and final Hamiltonian, one can optimize the fidelity under  influence of the noise within the nonadiabatic regime. 
\section{General formalism and results}
The dynamics of the quantum system is described by:
\beq
\label{eq:master_equation}
\frac{d \rop}{dt}=-\frac{i}{\hbar}[\Hop(t),\rop]+\mathcal{D}\rop,
\eeq
with 
\beq
\label{eq:dissipator}
\mathcal{D}\rop=- \frac{\gamma}{\hbar^2} [\Hop(t),[\Hop(t),\rop]], \quad \gamma>0 .
\eeq 
Here $\Hop(t)$ is the total Hamiltonian of the system including the control term, and 
the noise term is described by Eq. (\ref{eq:dissipator}). 
This term may result either from weak and continuous measuring (monitoring) the Hamiltonian \cite{diosi88,wisemanbook} or  from noise in the action,
caused for example by an error in scheduling the control Hamiltonians. 
Equation (\ref{eq:master_equation}) was derived and studied in the past for time independent Hamiltonians using a Poisson model and is sometimes referred to as intrinsic decoherence \cite{milburn91}. Here, we give a sketch of the derivation of Eq. (\ref{eq:master_equation}) for a time dependent Hamiltonian. 
We define the infinitesimal change in the action as,
\beq
\Delta s=\mathcal{H}(t)(dt+\sqrt{\gamma}d\xi), 
\eeq
where $\mathcal{H}(t)=-(i/\hbar)[\Hop(t),\cdot]$, and $d\xi$ is a stochastic increment satisfying, $\Mean d\xi=0$ and $(d\xi)^2= dt$, being $\Mean$ the stochastic mean and $\gamma$ the scale of the noise strength in units of time.
Henceforth, we define $\hbar=1$.
The infinitesimal change in the stochastic state $\sop$ is,
\beq
\label{eq:stoch_state}
\sop+d\sop=\exp(\Delta s)\sop.
\eeq
Expanding the rhs of Eq. (\ref{eq:stoch_state}) into a  series, keeping terms of order $dt$ and taking the stochastic mean, we obtain Eq. (\ref{eq:master_equation}) for the averaged state $\rop=\Mean \sop$. 

For a time independent Hamiltonian, the superoperator $\mathcal{D}$ leads to pure dephasing.
This implies that if the initial state $\rop(0)$ of the system is an eigenstate of the Hamiltonian and is diagonal in this basis, then the state is invariant during the dynamics.
Once the Hamiltonian  is time dependent, the state is no longer invariant and coherence will be created. 
This coherence will decohere by the noise leading to dissipation and a decline in the fidelity.
\par
\begin{figure}[t]
\center{\includegraphics[width=8.6cm]{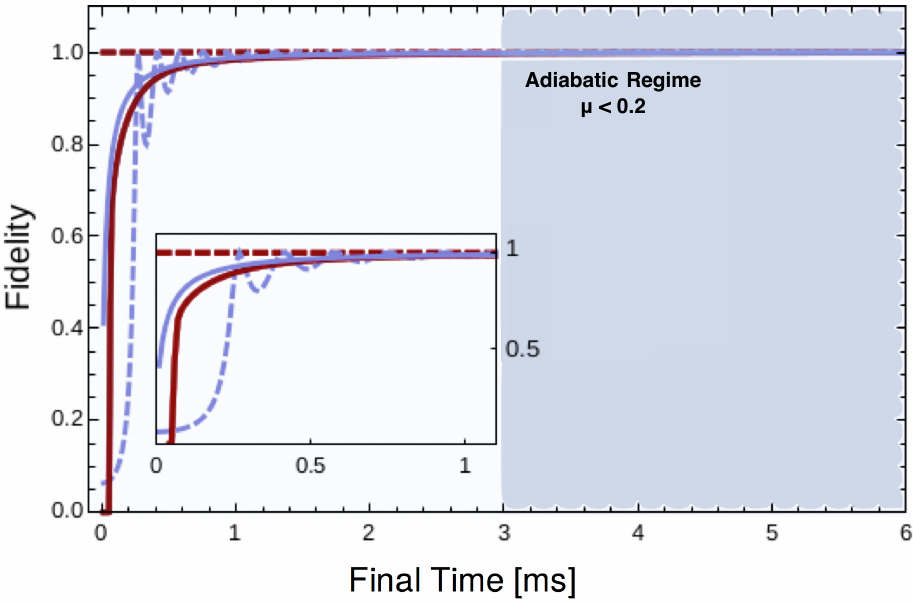}}
\caption{Fidelity of the harmonic oscillator for different final times  $t_f$. In red (dark grey) SP and in blue (light grey)  
adiabatic protocol. 
(Dashed lines) fidelity without noise and (solid lines) noise included in the dynamics.
The dark shaded area indicates the adiabatic regime for both protocols. Here, $\omega(0)= 2.5 \text{MHz}$, $\omega(t_f)= 2.5 \text{KHz}$ and $\gamma=0.8\text{ms}$.}
\label{fig:fig1}
\end{figure}

This behavior is predicted by looking at the equations of motion for the density matrix elements in the instantaneous eigenbasis of the Hamiltonian, $\rho_{kl}\equiv \bra{k}\rho \ket{l}$,  where $\lbrace \ket{k}\rbrace\equiv \lbrace\ket{k(t)}\rbrace$.
The coherences given by the off diagonal terms satisfy,
\beqar
\label{eq:coher_dynamics}
\dot{\rho}_{kl} & = & \left(i\epsilon(t)-\gamma \Delta E_{kl}^2(t)\right)\rho_{kl}+\braket{ k}{\partial_t l}\left(\rho_{kk}-\rho_{ll} \right)\\ \nonumber
& + & \sum_{n\neq k,l} \braket{\partial_t k}{n} \rho_{nl}+  \braket{n}{\partial_t l}  \rho_{kn} 
\quad \text{for} \quad k \neq l, 
\eeqar 
where  $\Delta E_{kn}(t)\equiv E_k(t)-E_n(t)$ is the gap between the eigenvalues of the Hamiltonian with  $n\neq k$ and the phase $\epsilon(t)$ is given by the sum of Berry and dynamical phases \cite{berry_phase},
\beq
\epsilon(t)=i\left(\braket{k}{\partial_t k}-\braket{l}{\partial_t l}\right)-\Delta E_{kl}(t), \quad \epsilon(t) \in \mathbb{R}.
\eeq
The diagonal terms satisfy
\beq
\label{eq:pop}
\dot{\rho}_{kk}=2\sum_{n\neq k}\Re\left(\rho_{nk} \braket{\partial_t k}{n} \right).
\eeq
The change in the population depends on the coherence $\rho_{nk}$ and on the overlap $\braket{\partial_t k}{n} = \bra{k}\partial_t \Hop(t) \ket{n}/\Delta E_{kn}(t)$.  
This term  does not depend explicitly on the noise implying that the noise influence enters only through the coherence.
The dynamics of the coherence, Eq. (\ref{eq:coher_dynamics}), is more involved. 
The first term on the rhs has both pure imaginary, $i\epsilon(t)$,  and real, $\gamma \Delta E_{kl}^2$, contributions.
The real part  $\Gamma(t)\equiv \gamma \Delta E_{kl}^2(t)>0$ is responsible for dephasing and is the direct consequence of the noise.
The dephasing rate $\Gamma(t)$ which is now time dependent is proportional to the square of the instantaneous eigenvalue separation and to $\gamma$.
Thus, changing the dephasing rate can be achieved by either manipulating the eigenvalues of the Hamiltonian or by controlling $\gamma$.   
The second term in the equation indicates that change in coherence is also proportional to the population difference between the two connecting levels.
The last term in Eq. (\ref{eq:coher_dynamics}) accounts for transitions from other off-diagonal elements. 
If the protocol changing the Hamiltonian is done adiabatically, i.e., the change is sufficiently slow, then  neither coherence nor excitation are generated during the protocol and the state follows the instantaneous eigenstate of the Hamiltonian with complete fidelity as can be observed in Fig. \ref{fig:fig1}.
In the nonadiabatic regime, $\text{max}\left\lbrace |\braket{k}{\partial_t l}|,t \in [0,t_f]\right\rbrace$ becomes large, which  then creates coherence and excitation.
%
%
%
%
%
%
%
%
%
%
%
\subsection{The harmonic oscillator}
Figure \ref{fig:fig1} shows the fidelity of a particle (or the center of mass of a particle cluster) with mass (or reduced mass) $m$ in a time dependent harmonic trap for various  final times $t_f$ using two different protocols to modify the trap frequency $\omega(t)$.
The Hamiltonian of the particle is given by,
\beq
\Hop(t)=\frac{1}{2m}\pop^2+
\frac{m\omega^2(t)}{2}\qop^2,
\eeq
with $[\qop,\pop]=i$.
The system is initially in the ground state of the Hamiltonian with frequency $\omega(0)=2.5 \text{MHz}$, and the target state is the corresponding ground state for $\omega(t_f)=2.5 \text{KHz}$.
The two different protocols for driving the Hamiltonian are a
shortcut to adiabaticity protocol (SP) \cite{sta_prl1} in red and 
adiabatic with constant $\mu$ in blue, where $\mu$ is the adiabatic parameter for closed systems \citep{aharonov87},
\beq
\label{eq:mu}
\mu=\sum_{l\neq k} \frac{\vert\bra{k}\partial_t \Hop \ket{l}\vert}{\Delta E_{kl}^2}
\eeq
For $\mu\ll 1$ the process is  performed in the adiabatic limit.
In \cite{lidar05} the conditions for adiabaticity for open quantum systems is derived by demanding that the Hilbert-Schmidt space can be decomposed
into decoupled Lindblad-Jordan eigenspaces. 
Since in this work the Lindblad operator is the Hamiltonian itself and the initial state is an eigenstate of $\Hop(t)$,  more insight can be gained by considering Eqs. (\ref{eq:coher_dynamics}) and (\ref{eq:pop}).
In this case adiabaticity implies $\dot{\rho}_{kk} \rightarrow 0$, which can be achieved either if $\mu \ll 1$ or $\rho_{nk}\rightarrow 0$. We will show that the latter can be achieved by manipulating $\Gamma(t)$.
\par
For the harmonic oscillator, $\mu(t)=\vert\dot{\omega}(t)\vert/\omega^2(t)$ \cite{kosloff17entropy}, which implies that for constant $\mu$, the Hamiltonian changes in time according to, 
\beq
\omega_{\mu}(t)=\frac{
\omega(0)\omega(t_f)t_f}{\omega(t_f)t_f-\left(\omega(t_f)-\omega(0)\right)t}.
\eeq
The fidelity, 
\beq
\mathcal{F}\defeq \text{tr}\sqrt{\sqrt{\rop(t_f)}\rop_{tar}\sqrt{\rop(t_f)}}
\eeq
measures the overlap between the final state and the target state $\rop_{tar}$. 
For Gaussian states the fidelity  is calculated according to \cite{banchi15}. 
We remark here that although a Gaussian state is not invariant under the dynamics generated by $\mathcal{D}$, Eq. (11) of \cite{banchi15} still provides a good measure for the fidelity of slightly perturbed Gaussian states.
\par
This example demonstrates that noise in the action is sensitive for short time operations. 
For both protocols,  as the dynamics become more adiabatic (i.e., small $\mu$), the fidelity asymptotically approaches one.
In the SP $\mu=\text{max}\lbrace\mu(t),t\in[0,t_f]\rbrace$, while for the adiabatic protocol $\mu$ is constant in each trajectory in the domain $[0,t_f]$.
%
\par
The above example illustrates  that in order to suppress the noise it is better to work in the adiabatic regime. 
Nevertheless, longer protocols are more sensitive to other noise sources including thermal and amplitude noises.
\subsection{The two level system}
In the next example we concentrate on the nonadiabatic regime (i.e., large $\mu$) and show how the noise can be manipulated to  improve fidelity.   
Our  study is the full population transfer of a two level system. 
This model has been studied extensively, see  \cite{muga12,demirplak03,berry09,muga12b,rabitz08} and references therein. 
It is important to note that the insight gained from these examples applies to the general control problems influenced by action noise.    
We assume a Hamiltonian of the form
\beq
\Hop(t)=\frac{ \Delta(t)}{2}\sop_z + \frac{\Omega(t)}{2}  \sop_x,
\eeq
where $\Delta(t)$ and  $\Omega(t)$ are real, time-dependent  functions resulting from an interaction with some external field, and $\sop_z$ and $\sop_x$ are the Pauli matrices. 
Initially the system is set to the ground state, $\rop=\ketbra{0}{0}$, with the initial Hamiltonian  corresponding to $\Delta(0)=\Delta_0$ and $\Omega(0)=0$, and a target of the exited state  $\rop=\ketbra{1}{1}$ with the final Hamiltonian $\Delta(t_f)=-\Delta_0$ and $\Omega(t_f)=0$.
As in the previous example, the dynamics is governed by Eq. (\ref{eq:master_equation}) and two protocols are considered: adiabatic (with constant $\mu$) and SP.
\par
\begin{figure}[t]
\center{\includegraphics[width=8.cm]{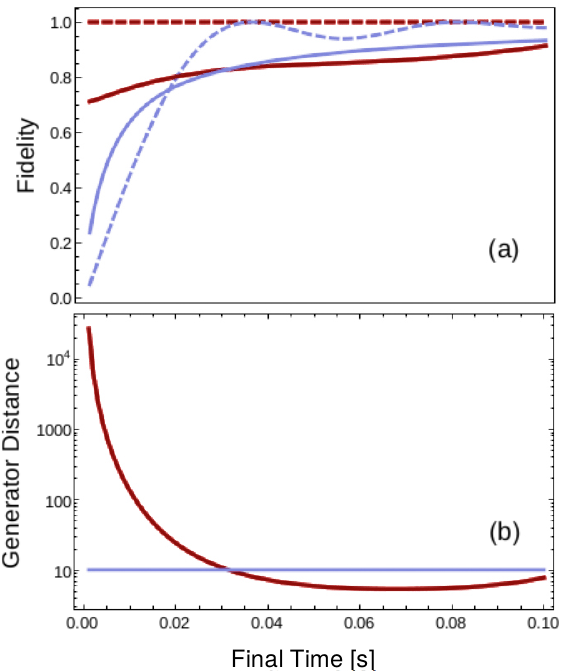}}
\caption{ (a) Fidelity of the two protocols in the nonadiabtic regime for different final times $t_f$. In red (dark grey) SP and in blue (light grey) ARP. 
(Dashed lines) ideal dynamics without noise and (solid lines) noise included in the dynamics.
(b) Generator distance $\mathcal{G_D}$ as a function of $t_f$ on a logarithmic scale. In red (dark grey) SP and in blue (light grey) ARP.  
Here: $\Delta_0=150 \text{Hz}$ and $\gamma=0.01s$. }
\label{fig:fig2}
\end{figure}

 
The adiabatic parameter is given by,
\beq
\mu = \frac{\vert\Delta(t)\dot{\Omega}(t)-\Omega(t)\dot{\Delta}(t)\vert}{2\left(\Delta^2(t)+\Omega^2(t) \right)^{3/2}}.
\eeq
It can be easily shown that the protocol for population inversion with constant $\mu$ satisfies,
\beq
\label{eq:ARP_protocol}
\Delta_{\mu}(t)=\Delta_0 \cos\left(\tfrac{\pi +2\pi n}{t_f} t\right),\quad \Omega_{\mu}(t)=\Delta_0 \sin\left(\tfrac{\pi +2\pi n}{t_f} t\right),
\eeq
with $n\in \mathbb{Z} $.
This protocol is also known as the adiabatic rapid  passage (ARP) \cite{metcalf07}, as full population inversion can be achieved in specific times within the nonadiabatic regime, see Fig. \ref{fig:fig2}. 
In this case $\mu=\tfrac{\pi}{2\Delta_0t_f}$. 
The SP is calculated according to \cite{inva_berry}.
%
%
%
%
%
%
%
%
%
%
\par
As in the example of the harmonic oscillator, we see that for both protocols the fidelity is increased with greater $t_f$,  see Fig. \ref{fig:fig2}.
In order to evaluate the significance of the noise as the protocol changes in time without having knowledge on the state, we define the generator distance $\mathcal{G_D}$ as the distance between the generator norm of the dynamics subject to noise and the ideal dynamics without noise, 
\beq
\mathcal{G_D} \defeq \frac{1}{t_f}\int^{t_f}_0 \Vert \mathcal{H}(t) +\mathcal{D}(t) \Vert - \Vert \mathcal{H}(t) \Vert dt.
\eeq
Where $\Vert\mathcal{B}  \Vert \defeq \text{max}\sqrt{\text{eig} \left( \mathcal{B}^{\dagger}\mathcal{B}\right)} $ is the spectral norm which is the largest singular value of $\mathcal{B}\in \mathbb{C}^{N\times N}$.  
As $\mathcal{G_D}$ increases, the noise becomes more significant along the trajectory.
\par
Figure \ref{fig:fig2}(b) presents the generator distance for the two protocols, SP and ARP. 
Shorter protocol times generate larger values of $\mathcal{G_D}$ for the SP (red line), indicating that operations become less sensitive to noise as the $t_f$ grows within the nonadiabatic regime.
For long times, in the adiabatic regime (not shown in the figure), $\mathcal{G_D}$ will begin to grow again. 
Nevertheless, in this regime no coherence will be created, thus the noise will not influence the dynamics.
This implies that the amount of coherence generated is crucial in determining the sensitivity of the control protocol to noise.
For shorter times, more coherence is generated, amplifying the sensitivity to noise. 
In the ARP, $\mathcal{G_D}$ is constant for different final times  $t_f$ as indicated by the blue line of Fig. \ref{fig:fig2}(b).  
This results from the fact that the instantaneous eigenvalues of the ARP Hamiltonian are time independent. 
\par
\begin{figure}[t]
\center{\includegraphics[width=8.6cm]{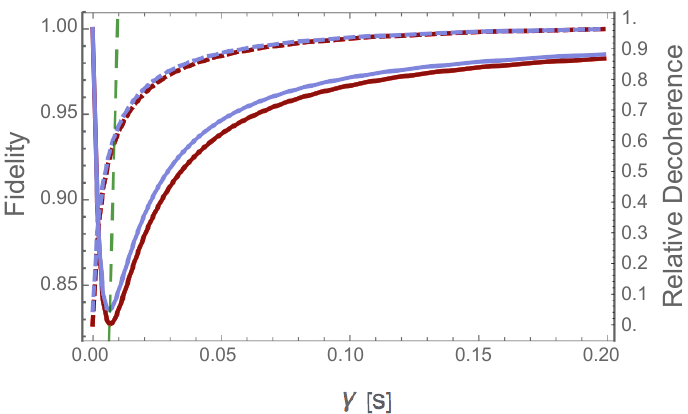}}
\caption{ (Solid lines) Fidelity of the two protocols as function of $\gamma$. 
(Dashed lines) Relative decoherence $\mathcal{C_R}$.
In red (dark grey) SP and in blue (light grey) ARP, the (dashed-green line) indicates the transition point.
Here:   $\Delta_0=150 \text{Hz}$ and $t_f=3.464\pi /(2\Delta_0)$ corresponding to the first oscillation pick in the fidelity of the ARP.}
\label{fig:fig3}
\end{figure}

 
As discussed above, merely analyzing the generator distance does not provide sufficient information about the effect of the noise on the dynamics.
To obtain the complementary picture, we evaluate the relative dissipated coherence along a certain process.
We define the average coherence generated along a trajectory as
\beq
\bar{\mathcal{C}}=\frac{1}{t_f}\int_0^{t_f}\mathcal{C}(t)dt,
\eeq 
 where $\mathcal{C}(t)=2\sum_{i\neq j}|\rop_{i,j}(t)|$ is the $l_1$ norm \cite{plenio14} of the off diagonal terms of the density operator in the instantaneous eigenbasis of the Hamiltonian.
The average dissipated coherence is evaluated as the difference between the average coherence of dynamics not including noise (ideal) $\bar{\mathcal{C}}_{id}$, and including noise $\bar{\mathcal{C}}_{n}$.      
We can then define the relative decoherence as the ratio between the dissipated and the ideal coherence
\beq
\mathcal{C_R} = \frac{\bar{\mathcal{C}}_{id}-\bar{\mathcal{C}}_{n}}{\bar{\mathcal{C}}_{id}}.
\eeq
This magnitude is bounded between zero and one. 
When $\mathcal{C_R}= 1$ it implies that all the coherence generated is on average decohered due to the noise.
In Fig. \ref{fig:fig3} we observe how the fidelity and the relative decoherence behave by varying $\gamma$ for fixed $t_f$.
Since $\mathcal{D}$ is linear in $\gamma$ the generator distance $\mathcal{G_D}$ monotonically increases with $\gamma$. 
The final time is fixed at $t_f=\frac{3.464\pi}{2\Delta_0}$  which corresponds to the first  peak of the ARP oscillation.
For the protocol Eq. (\ref{eq:ARP_protocol}), we
 find that the first peak is solely determined by $\Delta_0$.
The figure shows a transition point that as the noise gets stronger, the relative decoherence and the fidelity increase.
\par
This behavior can be understood in the following way: 
to complete the control protocol accurately in the nonadiabatic regime, a specific amount of coherence is generated along the trajectory.
When $\gamma =0$, no coherence is dissipated, i.e. $\mathcal{C_R}=0$. 
As $\gamma$ slightly deviates from zero, the relative decoherence $\mathcal{C_R}$ grows significantly and the fidelity declines. 
The SP can be executed in a shorter time than presented in Fig. \ref{fig:fig3} without changing $\Delta_0$.
Yet, for shorter  protocols, the change in both $\mathcal{C_R}$ and the fidelity will be more pronounced.
\par
In Fig. \ref{fig:fig3} when $\mathcal{C_R}\simeq 0.6$, the fidelity begins to grow with $\gamma$.
As $\gamma \rightarrow \infty$ the noise projects the state of the system on the instantaneous eigenstate of the Hamiltonian to match the target state with a fidelity of one. 
In this limit the process can be thought as a quantum Zeno effect where the instantaneous Hamiltonian is strongly monitored. 
%
%
%
%
%
%
%
%
%
\par  
A given experimental setup will not necessarily be able to fully control the parameter $\gamma$.
To achieve control over the action noise, the SP offer additional degrees of freedom which are absent in the ARP. 
That is, for a given initial and final Hamiltonian we can optimize the protocol of the SP \cite{oct_exp,oct_trans}, unlike the fully determined nature of the ARP.
To  control the dephasing rate $\Gamma(t)$ we define the parameter
\beq
\mathcal{M}=\sum_{ k\neq l}\frac{1}{t_f}\int_0^{t_f}\Delta E_{kl}^2(t)dt.
\eeq
Optimization of the protocol is carried out by constructing the Hamiltonian under a constraint on $\mathcal{M}$. 
One way to achieve this is by adopting the dynamical invariant method in which there are many possible Hamiltonians leading to the SP \cite{sta_prl1}. 
By imposing constraints on the protocol, the space of possible Hamiltonians is reduced. 
A large parameter $\mathcal{M}$ should be chosen in order to increase the relaxation rate, which projects the state onto the instantaneous eigenstates of the noise operator.
In other words, increasing the gap between the instantaneous eigenvalues of the Hamiltonian  will intensify the noise and eventually lead to higher fidelity.
\par
\begin{figure}[t]
\center{\includegraphics[width=8.6cm]{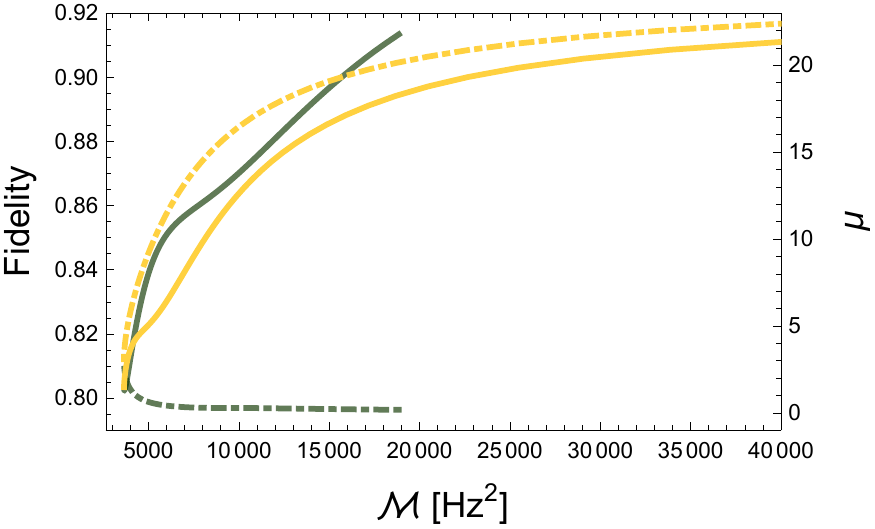}}
\caption{The green (dark grey) and yellow (light grey) lines correspond to two behaviors of the shortcut protocols.
Each point on these branches corresponds to a different protocol with different $\mathcal{M}$. 
The solid lines corresponds to the fidelity and the dashed line to the parameter $\mu$.
Here: $\Delta_0=150\text{Hz}$, $t_f=0.1s$ and $\gamma=0.01s.$
}
\label{fig:fig4}
\end{figure}

In Fig. \ref{fig:fig4} we plot the fidelity (solid lines) while modifying $\mathcal{M}$ for the dynamical invariant for the SPs. 
For each of these protocols the initial and final Hamiltonians are common.
In the intermediate times the protocols are constructed with an additional degree of freedom which enables  control over the parameter $\mathcal{M}$.
Modifying $\mathcal{M}$  also modifies the instantaneous eigenstates and thus changes $\mu$ and the amount of coherence and excitation along the trajectory. 
In the figure we observe two different behaviors, plotted in green and purple lines.
The  green line corresponds to increasing $\mathcal{M}$ while making the protocol more adiabatic, i.e. decreasing $\mu$. 
In this case Eq. (\ref{eq:mu}) is governed by the denominator. 
The fidelity is improved as a consequence of less coherence and excitation being  generated  and the increase of the dephasing rate $\Gamma(t)$.
\par
Since $\mathcal{M}$ is an averaged  quantity, a certain $\mathcal{M}$ may correspond to several different realizations of the SPs. 
The purple line represents a behavior where increasing $\mathcal{M}$ causes the process to be less adiabatic, i.e. grater $\mu$.
In this case the numerator in Eq. (\ref{eq:mu}) grows faster than the denominator.
Although $\mu$ increases along this branch, the large rate $\Gamma(t)$ guarantees that the state will closely follow the instantaneous  eigenstate of the Hamiltonian.   
The fidelity plotted in the purple solid line will continue to grow as $\mathcal{M}$ grows (although this regime is not plotted in the figure).
%
%
%
%
%
%
%
\par
\section{Discussion}
A recent publication by Kiely et al. on the effect of Poisson noise on adiabatic quantum control \cite{muga17} supports our claim that noise can improve the adiabatic  following for initial diagonal states $\rop(0)$, also in the weak noise limit (i.e. small $\gamma$).
While in their study the authors concentrated on adiabatic protocols, in the present study we focused on SP which are preferable for suppressing thermal noise.

We find that in the presence of action noise there is a trade-off between completing the SP in shorter times and the fidelity. 
Unlike noise induced by the environment, action noise becomes more disturbing as the protocol time is reduced.
This effect can be traced to a larger amount of coherence generated and bigger $\mathcal{G_D}$ for shorter protocols.
We expect a similar result for the Poisson model studied in \cite{muga17}.
We further showed that the SP admits higher controlability over the noise compared to ARP, as the Hamiltonian's eigenvalues and eigenvectors can be manipulated along the process.
By doing so, both $\mu$ and the relaxation rate $\Gamma(t)$ are controlled.
That is, for a given $\gamma$, finite time $t_f$ and initial and final Hamiltonians, we can manipulate the effect of the noise in the intermediate times and achieve higher fidelity with respect to the target state.
To attain control over initial states that are in a superposition in the basis of the initial Hamiltonian, contrary to the scheme introduced above,  we suggest  to minimize $\mathcal{M}$ and thus minimize the rate $\Gamma(t)$. This will be considered in a future study.    

In \cite{levy12} it was suggested that noise can be used as a resource for constructing a quantum absorption refrigerator, and a recent experimental work on trapped ions have supported this claim \cite{maslennikov17}.
In \cite{koch15b} back-flow of amplitude and phase from the environment into the system was exploited to carry out quantum control tasks that couldn't be realized solely by unitary protocols. 
An experiment using the quantum Zeno effect to control a qubit was recently reported \cite{hacohen17}.  
Thus, for some applications it is interesting to exploit rather than suppress the noise in order to manipulate quantum systems in experimental designs.
%
%
%
%
\section{acknowledgements}
We acknowledge J. G. Muga for fruitful discussions and for drawing our attention to his related work \cite{muga17}. We are grateful to L. McCaslin for useful comments. 
This work was funded by the US Army Research Office under Contract W911NF-15-1-0250 and the Basque Government (Grant No. IT986-16), MINECO/FEDER,UE (Grants FIS2015-67161-P and FIS2015-70856-P), and QUITEMAD+CM S2013-ICE2801.


\end{document}